\documentclass[cite,figures]{epl}
\usepackage{psfig}

\newcommand{\putfiga}
{
\begin{figure}[bt]
\centerline{%\hspace*{0.15cm}
\psfig{file=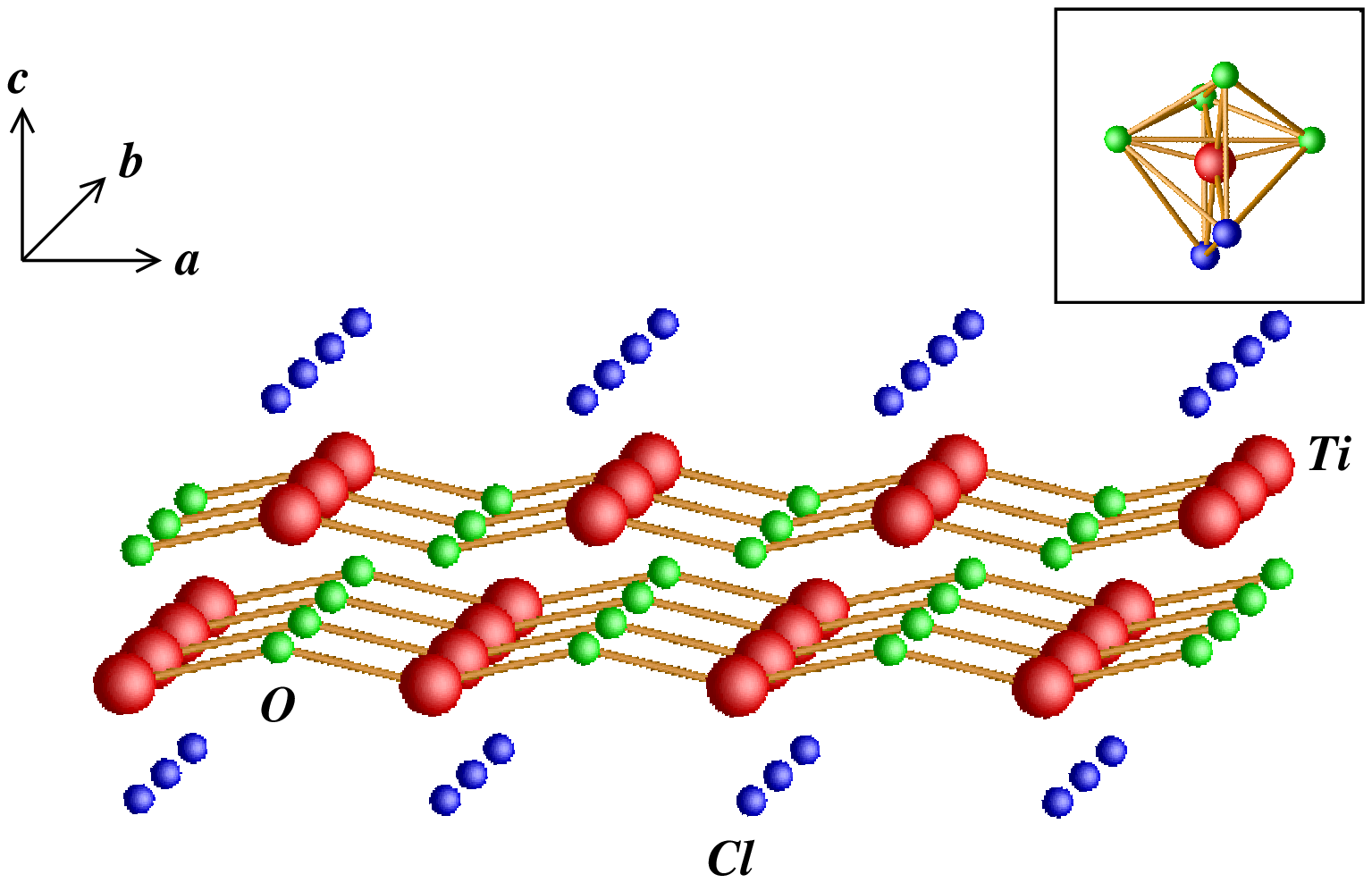,width=0.47\textwidth}
\hspace{1cm}
\psfig{file=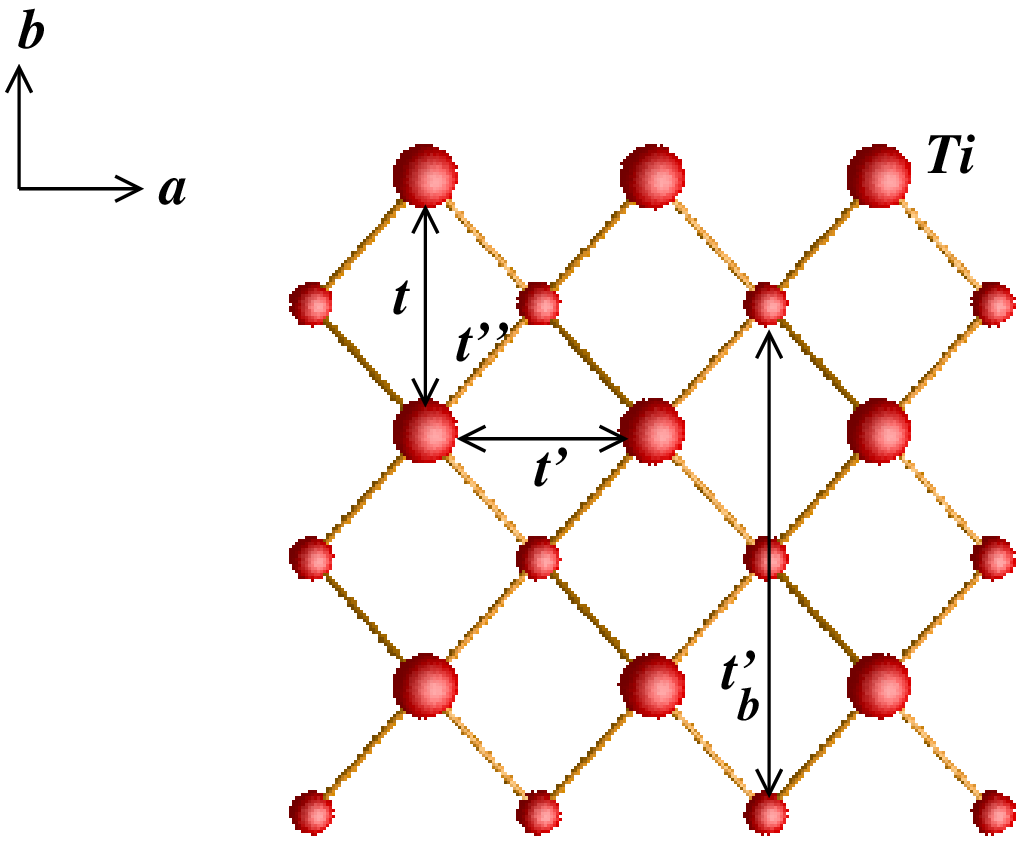,width=0.38\textwidth,angle=0}}
\vspace*{-0.4cm}
\caption{\label{structure} Crystal structure of TiOCl.   Left panel:
 layered structure of the compound and the distorted
octahedral Ti environment (see inset). Right panel: 
network of Ti atoms projected in $ab$ plane with the various hopping
paths. Seen in the figure are two layers of Ti atoms. The two different
sizes of the Ti atoms correspond to atoms belonging to two different layers.  }
\vspace{-.4cm}
\end{figure}
}

\newcommand{\putfigb}
{
\begin{figure}[tb]
\centerline{
\psfig{file=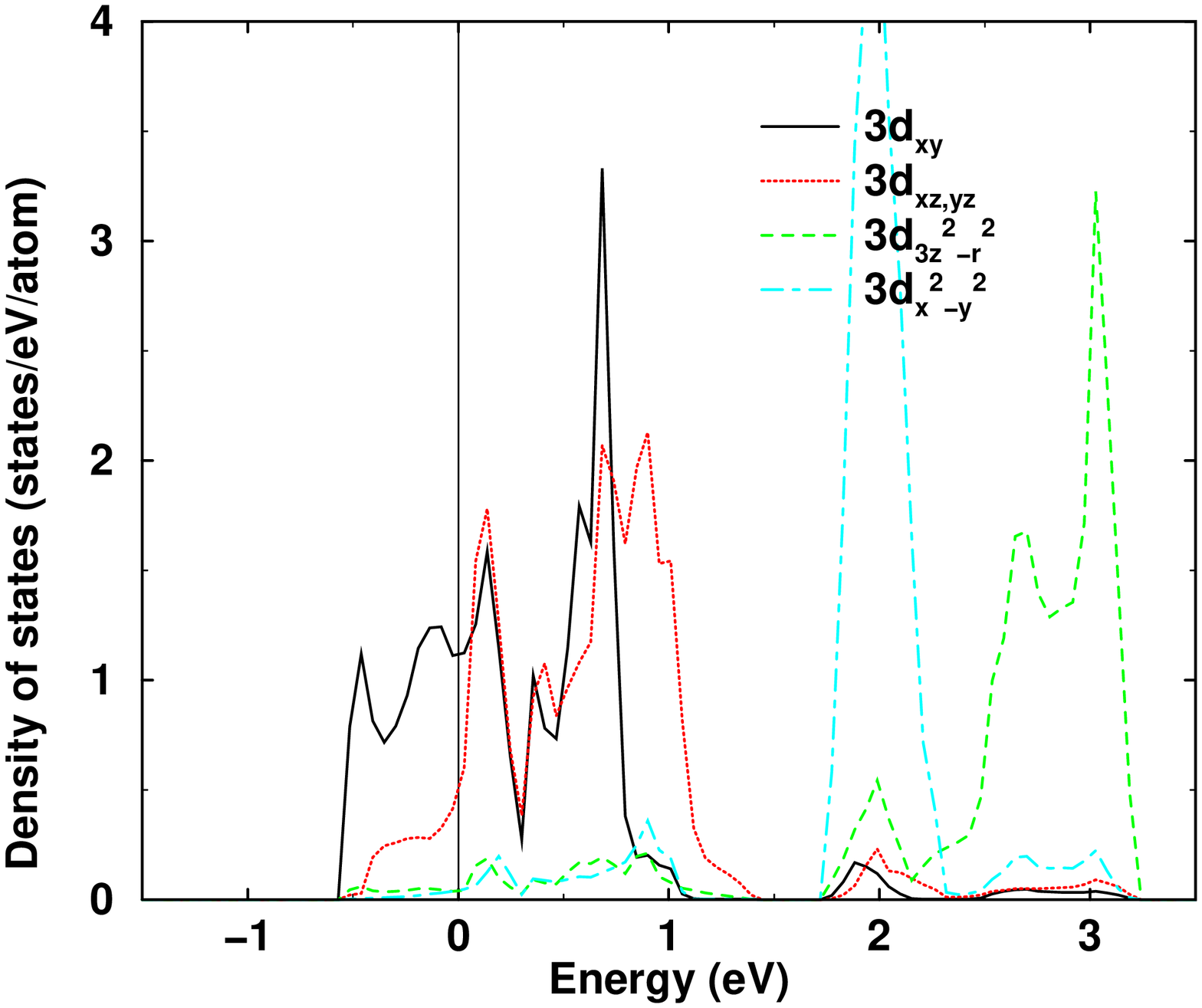,width=0.4\textwidth,angle=-90}
\hspace{.5cm}
\psfig{file=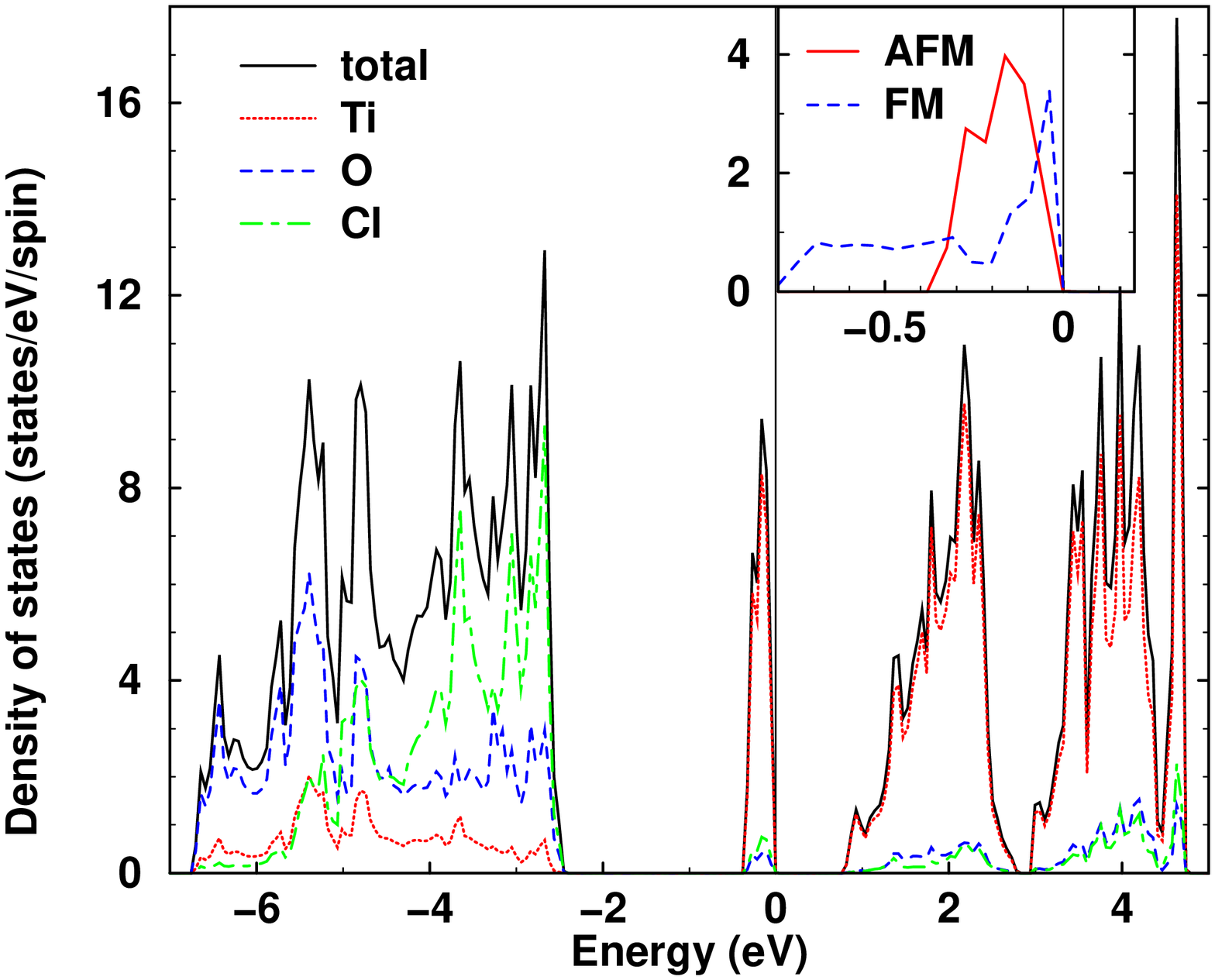,width=0.41\textwidth,angle=-90}}
\vspace*{-0.4cm}
\caption{\label{dos} FPLO-LDA-Density of states for TiOCl. Left panel:
Orbital resolved Ti-3d states for the non spin polarized case. The
3$d_{xy}$ orbital lies at lower energy than the
3$d_{xz,yz}$ states.  The $t_{2g}$-$e_g$ split is about 2 eV.  Right
panel: LDA+U results for AFM Ti spin arrangement along $b$.  The O-2$p$
and Cl-3$p$ states are mainly occupied but show sizable hybridization
with the Ti 3$d$ states. The narrow peak close to the Fermi level is
predominantly Ti-$3d_{xy}$.
In the inset the Ti $3d_{xy}$ states for the AFM
and FM configurations are compared.}
\vspace{-.4cm}
\end{figure}
}

\newcommand{\putfigc}
{
\begin{figure}[bt]
\vspace{-.55cm}
\centerline{
\psfig{file=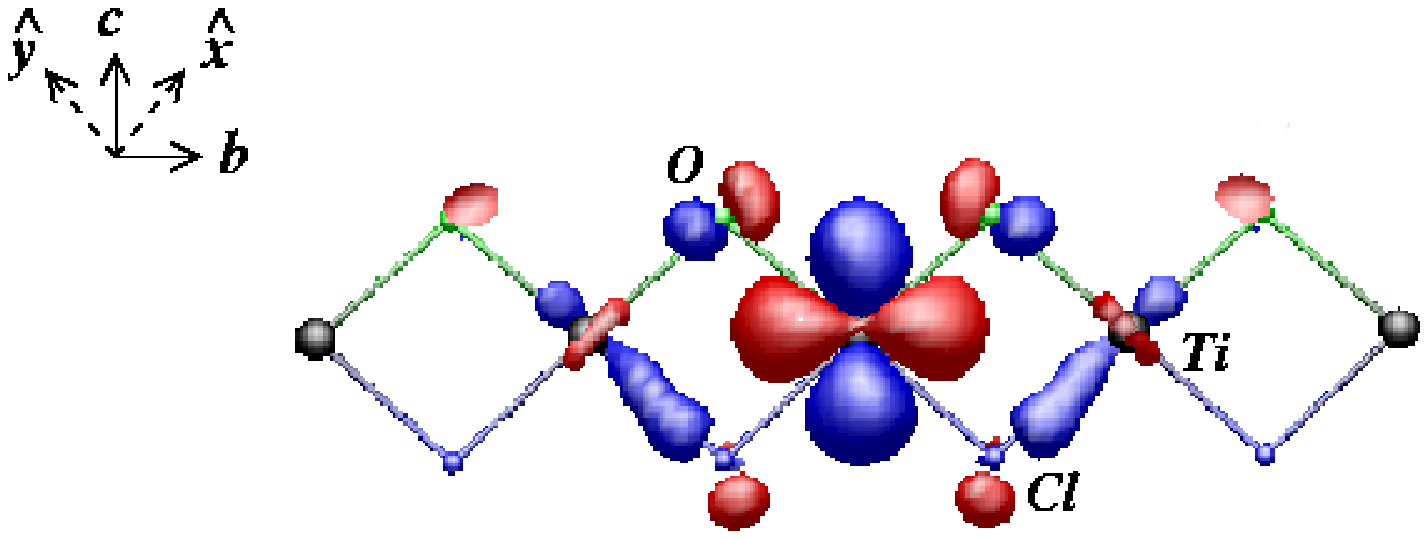,width=0.45\textwidth}}
\vspace*{-0.6cm}
\caption{\label{ED_df} Effective Ti-d$_{xy}$ orbital
(Wannier orbital) plotted in the $bc$ plane in the downfolded representation
where all the channels other than Ti-d$_{xy}$ have been integrated out.}
\vspace{-.4cm}
\end{figure}
}

\newcommand{\putfigd}
{
\begin{figure}[bt]
\vspace{-.3cm}
\centerline{
\psfig{file=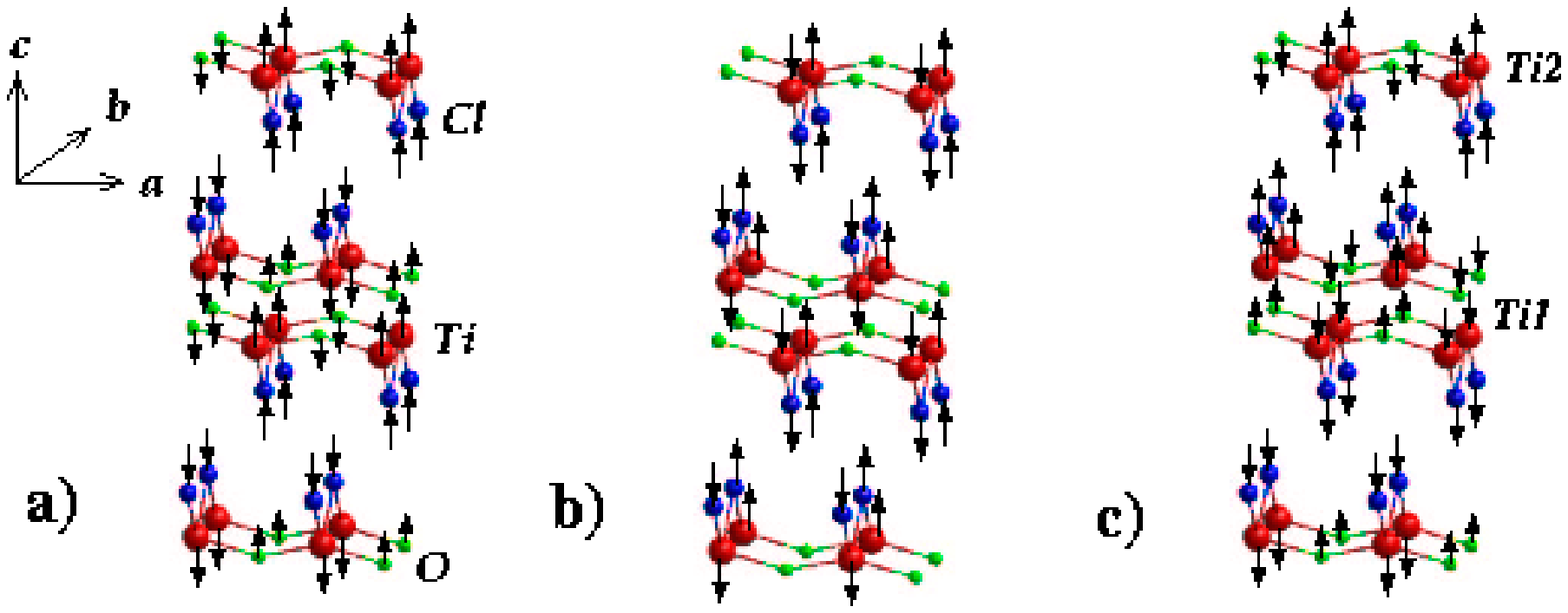,width=0.88\textwidth}}
\medskip

\caption{\label{distort}
 Ti-Cl in-phase distorted structures according to various
 allowed 
 phonon modes in TiOCl.
 a) Distortion with no doubling of the unit cell. Here
two different cases can be possible, one in which 
the Ti atoms belonging to two different layers within a given
bilayer are pushed towards each other as shown in the figure 
and another in which the Ti atoms are pulled apart 
(obtained by reversing the direction of arrows).
 b) Distortion with the unit cell doubled along $b$.
 c)  Distortion with the unit cell doubled along $c$. Here the 
 Ti  belonging to two different layers within the bilayers are
alternately pulled apart in one bilayer (Ti1 atoms) and pushed
towards each other
 in the next bilayer
(Ti2 atoms).
}
\vspace{-.4cm}
\end{figure}
}

\newcommand{\putfige}
{
\begin{figure}[tb]
\centerline{\hspace*{0.0cm}
\psfig{file=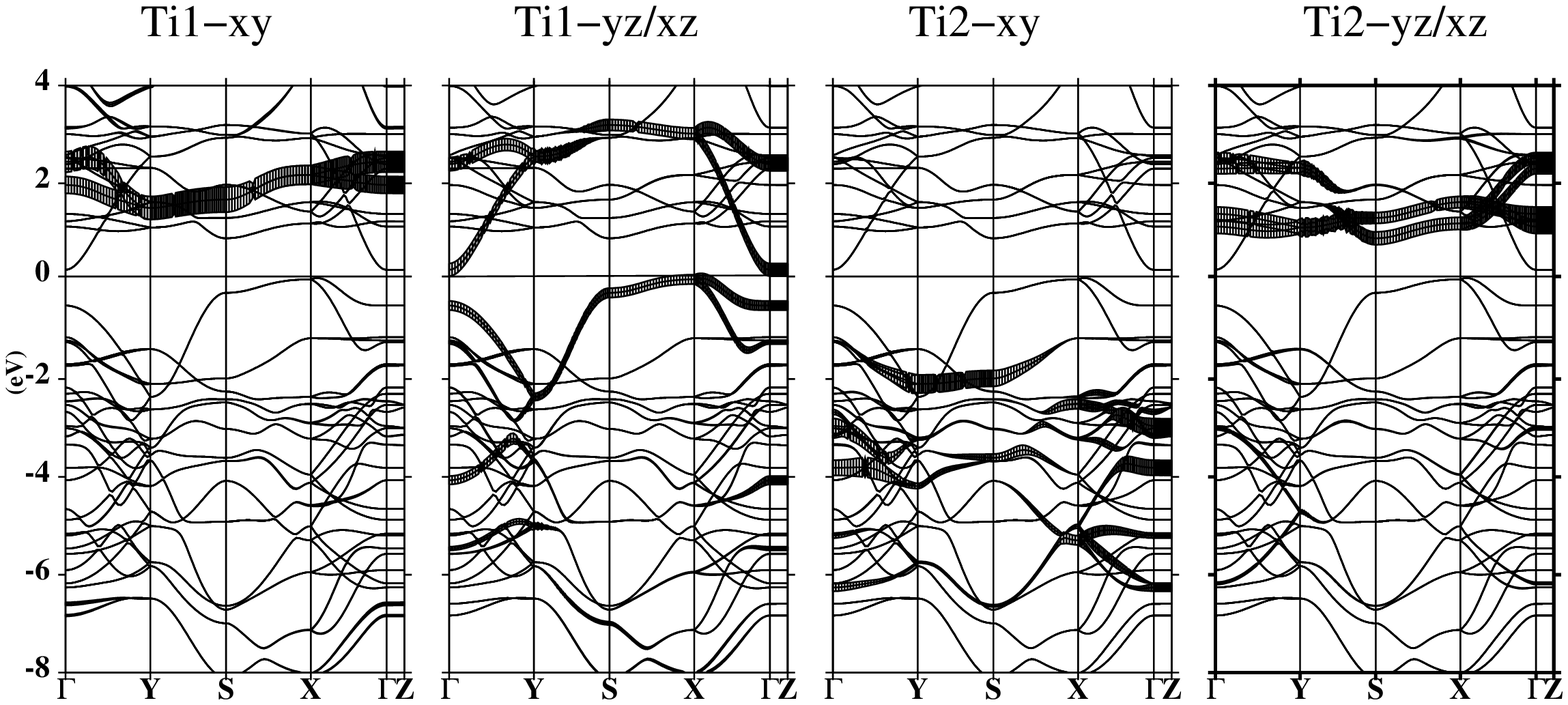,width=0.8\textwidth}}
\vspace*{-0.1cm}
\caption{\label{distorsion_bands} LMTO-LDA+U bandstructure 
in FM configuration
 for the distorted structure shown in Fig.~\ref{distort} c).
Only the bandstructure corresponding to the majority spin bands
is shown, since for the minority spin the $d$-bands are
completely unoccupied.
The fatness of the bands in each panel corresponds
to the weight of the indicated orbital in the wavefunction.
The first and second
panel show that the occupied orbitals for Ti1 are the 
near-doublet $d_{yz}$, $d_{xz}$ while $d_{xy}$ is higher in energy.
The third and fourth panel show that for  Ti2, $d_{xy}$ remains
the lowest-energy, occupied orbital. }
\vspace{-.6cm}
\end{figure}
}

\begin{document}
\title{ TiOCl, an orbital-ordered system?}
\shorttitle{ TiOCl, an orbital-ordered system?}

\author{T. Saha-Dasgupta\inst{1} \and Roser Valent\'\i\inst{2} \and
H. Rosner\inst{3} \and Claudius Gros\inst{4}}
\shortauthor{T. Saha-Dasgupta \etal}

\institute{\inst{1} S.N. Bose National Centre for Basic Sciences, JD
Block, Sector 3, Salt Lake City, Kolkata 700098, India\\
\inst{2}
Institut f\"ur Theoretische Physik, Universit\"at Frankfurt,
D-60054 Frankfurt, Germany\\
\inst{3}
Max-Planck Institute for Chemical Physics of Solids,
%Noethnitzer Str. 40,
D-01187 Dresden, Germany\\
\inst{4}
Theoretische Physik, Universt\"at des Saarlandes, D-66041 Saarbr\"ucken
Germany}
%\pacs{75.30.Gw, 75.10.Jm, 78.30.-j }

\maketitle
\vspace{-1cm}
\begin{abstract}
\footnotesize{
%{\it paper version: Nov 24 2003}\\ %date command does not work
We present first principles density functional calculations and
downfolding studies of the
electronic and magnetic properties of the layered quantum spin system
TiOCl.  We discuss explicitly the nature of the exchange paths
and  
the concept of {\it orbital ordering} in this material.  
An analysis of the electronic structure of slightly distorted
structures
according to various phonon modes allowed in this material suggests
that this system  may be subject to  orbital fluctuations driven
by the electron-phonon coupling.  

}
\end{abstract}
%%%%%%%%%%%%%%%%%%%%%%%%%%%%%%%%%%%%%%%%%%%%%%%%%%%%%%%%%%%%%%%%%%%
%%%%%%%%%%%%%%%%%%%%%%%%%%%%%%%%%%%%%%%%%%%%%%%%%%%%%%%%%%%%%%%%%%%
Spin=1/2 quantum spin systems involving early transition metal ions
with a 3d$^1$ configuration are being lately intensively studied since
they present a large variety of interesting  novel phenomena.
Examples of these materials are those containing V$^{4+}$ or Ti$^{3+}$
ions where one $d$ electron occupies one of the three $t_{2g}$
orbitals and competing spin, charge, and orbital-degrees of freedom
determine their properties.  For instance, NaV$_2$O$_5$ is a
mixed-valence system with V$^{4.5+}$ which undergoes at T=34K a phase
transition to a charge-ordered state (V$^{4+}$/V$^{5+}$) accompanied
by the opening of a spin-gap and a lattice distortion.  The origin of
this phase transition is not yet completely
settled\cite{review}.  
Another interesting system is LaTiO$_3$
where an unusual many body state with antiferromagnetic (AFM) long-range
order and orbital fluctuations has been observed\cite{Keimer_00} and
 discussed at length in relation to a possible quantum orbital liquid
phase\cite{Khaliullin_00}.  This interpretation though, is still
unclear in view of recent X-ray and neutron diffraction
data\cite{Cwick_03} as well as
the dynamical mean field theory (DMFT)
coupled with local density approximation (LDA)
calculations\cite{Pavarini_03,Craco_03} which support the existence of
non-degenerate $t_{2g}$ orbitals in LaTiO$_3$.

Here we want to focus our attention on the layered system TiOCl which
has recently raised a lot of discussion.  Theoretical considerations as
well as susceptibility studies\cite{Seidel_03}, NMR \cite{Imai_03},
ESR \cite{Kataev_03} and infrared and Raman spectroscopy
\cite{lemmens_03_2} have been reported.  Susceptibility measurements
show a kink at T$_{c2}$=94 K and an exponential drop at T$_{c1}$=66 K
indicating the opening of a spin gap\cite{Seidel_03}. This feature
which  seems to be accompanied by a doubling of the unit-cell along the
$b$-axis\cite{Imai_03}(see Fig.~\ref{structure}) has been originally
described as a spin-Peierls transition.  However, the observation of
large phonon anomalies\cite{lemmens_03_2} as well as
temperature-dependent g-factors and linewidths in ESR\cite{Kataev_03}
indicate that this is not a conventional spin-Peierls transition but
rather a transition driven by competing lattice, spin, and orbital
degrees of freedom with possible orbital fluctuations.

LDA+U calculations for a ferromagnetic (FM)
groundstate using the full-potential LMTO
method\cite{Andersen_75} suggest that the equilibrium state above $T_{c2}$
is described by one dimensional spin-1/2 chains running along the
$b$ axis where the spins are localized in  Ti $d_{xy}$
orbitals\cite{Seidel_03}.  
Since the three $t_{2g}$ orbitals are close in energy, a scenario 
with possible orbital ordering in this material has also been 
suggested\cite{Seidel_03,Imai_03}.  We shall investigate this proposal 
hereafter.

%%In this letter it is our purpose:
The purposes of our study in this letter are the following:
(i) to supplement Seidel {\it et al.}'s \cite{Seidel_03} work on TiOCl
for the crystal structure measured at high temperatures T $>$ T$_{c2}$
 by  performing a detailed study of the groundstate 
of the system in terms of various probable magnetic structures
(ii) to calculate explicitly hopping parameters between Ti ions and the 
nature of the exchange paths with the downfolding procedure\cite{newlmto}
(iii) finally and importantly, motivated by the signature of 
strong coupling between lattice and spin degrees of freedom\cite{Kataev_03,
lemmens_03_2}, to perform a frozen phonon study in 
order to understand the influence of the ion displacements on the 
electronic ground state.
Our results suggest that  though the high temperature phase above T $>$
T$_{c2}$ is certainly not an orbital ordered state, during the vibration of 
certain phonon modes, occupation of various $t_{2g}$ orbitals can change 
drastically and hence this system may be subject to  orbital
fluctuations driven by the electron-phonon coupling.  
 No such theoretical calculations have been attempted before and
may provide a useful guide in understanding this interesting material.

\putfiga
%\vspace{0.2cm}

{\it Crystal structure and ligand field}.-  The high-temperature 
phase of TiOCl crystallizes
in the $Pmmn$ space group and consists of bilayers
of Ti$^{3+}$ and O$^{2-}$ parallel to the $ab$ plane separated by
layers of Cl$^{-}$ (see Fig.~\ref{structure}).  The basic unit in this
structure, i.e. the TiCl$_2$O$_4$ octahedron (shown as inset in Fig.\
\ref{structure}) is  distorted with distances
d(Ti-O$_{apical}$)= 1.95 \AA, d(Ti-O$_{equatorial}$)= 2.25 \AA ,
d(Ti-Cl)= 2.32 \AA\ and the angle
$\alpha$(O$_{apical}$-Ti-O$_{apical}$)$\approx$ 153$^o$.  In a perfect
octahedral environment 
%when all ligand-ions are equally distant to
%the center and positioned on the main axes, 
the crystal field splits
the five degenerate $d$ orbitals into (in the local reference
frame) the low energy triplet $t_{2g}$=
($d_{xy}, d_{xz}, d_{yz}$) and the higher energy doublet $e_g$=
($d_{z^2}$, $d_{x^2-y^2}$).  In the case of TiOCl, a further splitting
of the $t_{2g}$ orbitals into a lower $d_{xy}$ and higher energy
$d_{xz}, d_{yz}$ occurs \cite{explain}.
These considerations  
suggest that the groundstate for
 TiOCl at T $>$ T$_{c2}$ should be described by the Ti
3$d_{xy}$ orbitals.

{\it Ab-initio calculations}.- We have performed density functional
calculations in the LDA,
generalized gradient approximation (GGA)\cite{Perdew_96} and in the LDA+U \cite{Anisimov_97}
approximation for this system using the linearized
muffin tin orbital (LMTO) method based on the Stuttgart TBLMTO-47
code\cite{Andersen_75} and the full potential minimum basis local orbital
code (FPLO)\cite{Koepernik_99}
We checked convergence of the band structure results
obtained with the LMTO scheme -where the downfolding procedure is
implemented- versus the results from the full potential scheme \cite{convergence}.

\putfigb

In Fig.~\ref{dos} we present the orbital-resolved density of states (DOS)
obtained in the FPLO scheme. The coordinate system has been chosen as
$\hat{z}$ = $a$, and $\hat{x}$ and $\hat{y}$ axes rotated 45$^{o}$
with respect to $b$ and $c$.
We observe that the degeneracy
between the $t_{2g}$ orbitals is already lifted by the ligand field
into $d_{xy}$ and the slightly higher in energy
  $d_{yz}$, $d_{xz}$\cite{ref_frame} (see
Fig.~\ref{dos}, left panel).
 TiOCl is an insulator although the LDA and GGA calculations
suggest metallic behavior due to the inadequate 
treatment of the Coulomb term in those approximations.  A
better description of the electronic correlation is obtained by the
LDA+U approach which takes into account the orbital dependence of the
Coulomb and exchange interactions \cite{Anisimov_97}.  Such a
calculation is carried out on a spin-polarized state.
The
earlier study by Seidel {\it et al.} \cite{Seidel_03} considered only
the FM arrangement of the Ti spins. In order to
investigate whether the nature of the magnetic configuration has any
influence on the orbital structure and in view of the possible antiferromagnetic
(AFM) coupling of Ti ions along the $b$-direction -as suggested by the AFM
Heisenberg chain model fitting the susceptibility data \cite{Seidel_03}-  we have 
  carried out further calculations 
where the Ti ions
are AFM aligned along $b$ and FM aligned along $a$ and $c$.  This
calculation
requires  the doubling of the unit cell along $b$
transforming the unit cell from orthorhombic to monoclinic.
In the LDA+U  an appropriate value for  $U$ and the
onsite exchange $J_o$ has to be taken.  Previous first principles
studies\cite{Seidel_03} considered a FM spin-polarized state with
$U=3.3 eV$ and $J_o=1 eV$.  We have chosen here the same $J_o$ and we
didn't
observe
significant changes  in the results by considering
a lower $J_o$ value ($ 0.3$ and $0.4$ eV)
which may be more appropriate for early transition metals like Ti.
U was taken to vary between 2-7 eV.
 The
%%%%%%%%%%%%%%%%%%%%%%%%%%%%%%%%%%%%%%%% converged LDA+U
results for both  FM and AFM structures show a gap opening between
occupied $d_{xy}$ orbitals and unoccupied $d_{xz}$ and $d_{yz}$
orbitals,  as seen in the LDA+U DOS shown on the right panel of
Fig.~\ref{dos}. We observe that the choice of a FM or AFM state
influences the total energy and the shape of the bands but not the
orbital nature of the groundstate. 
Independent of the choice of the $U$ value (see below), the AFM spin 
configuration was found to be lower in total energy compared to the FM state.
We have also considered a third possibility, namely Ti AFM aligned
along $a$ and FM along $b$ and along $c$. The LDA+U calculations show
again a bandstructure with a gap opening between occupied $d_{xy}$
orbitals and unoccupied $d_{xz}$ and $d_{yz}$ orbitals.
 
Our study corroborates the fact that the groundstate is indeed
described by chains of Ti ions along the $b$ direction occupying the
$d_{xy}$ orbitals as also observed in the interpretation of
ESR\cite{Kataev_03} and infrared spectroscopy\cite{lemmens_03_2}
results.  Consideration of other probable magnetic structures
as well as the study of the role played by U and $J_o$ in the LDA+U
 calculations
 along with the ligand field calculation of an isolated octahedron
prompts us to more definitely to conclude that at T$>$T$_{c2}$ TiOCl
cannot be described as an {\it orbital ordered} state, thereby leaving no
room for discussion on the possibility of the groundstate being described
by a {\it zig-zag} chain along $a$ formed by alternating approximately
degenerate $d_{xz}$ and $d_{yz}$  as was initially suggested as a
possible
scenario for TiOCl in Ref. \cite{Seidel_03,Imai_03}.

 We obtained the exchange constant $J$ between neighboring Ti along 
the $b$ direction by subtracting the LDA+U total energy in the
FPLO scheme between the AFM-coupled Ti spins along $b$  
and the totally FM spin configurations as
a function of U.  We found an almost linear dependency of $J$
with respect to
$1/U$ for $U$ values between 2 and 7 eV. For U $\sim$ 3.3eV the computed
AFM $J$ is about 750K, in good agreement with the
estimate for an AFM superexchange $J=4t^2/U$ and with susceptibility
measurements \cite{Seidel_03}.  This good agreement suggests that FM contributions to
the exchange play a minor role in this system.  For an analysis of
these results we proceed in the next section with a study of the
various interaction paths in this material.

{\it Downfolding: Effective Hamiltonians}.- In order to 
 provide an understanding 
of the interaction paths in TiOCl we have
applied the tight-binding-downfolding procedure implemented within the
framework of the improved version of the LMTO method, namely the NMTO
method \cite{newlmto}. The method can be used to
derive a few-orbital effective Hamiltonian $H_k$ 
starting from the full LDA or GGA Hamiltonian by folding down the 
inactive orbitals.
The Fourier transform of this few-orbital downfolded Hamiltonian
provides  {\it ab-initio} estimates of the hopping matrix elements of 
the corresponding tight-binding (TB) Hamiltonian defined in the effective orbital 
basis while the effective orbitals themselves provide the Wannier-like 
functions of the corresponding bands.
\begin{table}
\begin{center}
\begin{tabular}{|c||c|c|c|c|c|}
\hline
& Ti$xy$ & Ti$xy$+O$p$+Cl$p$ & Ti$xy$+O$p$ &Ti$xy$+Cl$p$  \\ \hline
t & -0.21 &  -0.26 & -0.16 & -0.31  \\
t$^{``}$ & 0.03 &  0.08 & 0.04 & 0.20  \\
t$^{'}$ & 0.04 &  -0.01 & 0.06 & -0.02  \\
t$^{'}_{b}$ & -0.03 &  -0.05 & -0.03 & -0.04  \\\hline
% &  &  &  & \\
\end{tabular}
\end{center}
\caption{ Predominant hopping integrals 
 (in eV) for TiOCl
at various levels of downfolding. Ti$xy$: all except the Ti$-xy$ channel
have been integrated out.  Ti$xy$+O$p$+Cl$p$: all except
 Ti$-xy$, O$-p$ and Cl$-p$ have been integrated out.
Ti$xy$+O$p$: all except   Ti$-xy$ and  O$-p$  have been 
integrated out. Ti$xy$+Cl$p$: all except  Ti$-xy$ and  Cl$-p$ 
 have been integrated out.
% everything other than the
%Ti$-xy$ channel has been integrated out, Ti$xy$+O$p$+Cl$p$: everything other 
%than the Ti$-xy$, O$-p$ and Cl$-p$ channels has been integrated out,
%Ti$xy$+O$p$: everything other 
%than Ti$-xy$ and  O$-p$ channels has been integrated out,
%Ti$xy$+Cl$p$: everything other 
%than Ti$-xy$ and  Cl$-p$ channels has been integrated out. 
The hopping paths are as shown in
Fig.\ \protect\ref{structure}.
}
\label{parameters}
\end{table}

In Table \ref{parameters} we present the hopping parameters (see 
 Fig.\ \ref{structure} right panel)  obtained from 
downfolding   the GGA
bands in TiOCl in order to define an effective Ti $d_{xy}$-only
Hamiltonian.
 We observe that the
predominant hopping ($t$) is between nearest neighbors Ti $d_{xy}$
along $b$  while
the hopping parameters between next
nearest neighbors along $b$ and along $a$ and $c$ directions are
small. The system therefore virtually behaves as one-dimensional. An
important point here is to investigate the nature of the (Ti
$d_{xy}$)-(Ti $d_{xy}$) interaction path.  The effective (Ti
$d_{xy}$)-(Ti $d_{xy}$) hopping obtained by downfolding all the
channels except Ti $d_{xy}$ (shown in the first column of Table
\ref{parameters}), involves the direct (Ti $d_{xy}$)- (Ti $d_{xy}$)
hopping renormalized by the contributions of the out-integrated O-$p$
and Cl-$p$ channels.  In order to quantify the contribution of O and
Cl to the various (Ti $d_{xy}$)-(Ti $d_{xy}$) interaction paths, we
considered the downfolding procedure by keeping active, in addition to
two Ti-$d_{xy}$ in the basis set, i) O-$p$ and Cl-$p$ orbitals
(Ti$xy$+O$p$+Cl$p$) ii) O-$p$ orbitals (Ti$xy$+O$p$) and iii)
Cl-$p$ orbitals (Ti$xy$+Cl$p$).  This method has been proven to be
very useful for determining the nature of the interaction paths in
the frustrated Cu$_2$Te$_2$O$_5$Br$_2$ and Cu$_2$Te$_2$O$_5$Cl$_2$
systems \cite{Valenti_03_t}.

The  TB parameters corresponding to the (Ti$xy$+O$p$+Cl$p$) set give an 
idea of the magnitude of the direct (Ti $d_{xy}$)-(Ti $d_{xy}$) overlap.  
On the other hand, downfolding to
(Ti$xy$+O$p$) and (Ti$xy$+Cl$p$)  accounts for
the (Ti $d_{xy}$)-(Ti $d_{xy}$) overlaps via Cl-$p$ and O-$p$ respectively
in addition to direct overlap. Comparing the numerical values of the
(Ti $d_{xy}$)-(Ti $d_{xy}$) hopping integral, $t$, at various levels of
downfolding we observe that while already the direct overlap is
considerable (in the range of energies that we are dealing here), the
contribution of O-$p$ and Cl-$p$ is not negligible.  This can be more
easily seen in  Fig.~\ref{ED_df} which shows  the
effective Ti d$_{xy}$ orbital in the downfolding calculation where the
channels other than Ti $d_{xy}$,  namely Cl-$p$ and the O-$p$, as well
as the other Ti-$d$ characters 
have been integrated out. The
contribution of the integrated-out orbitals appear as a {\it tail} of
the effective orbital with the {\it central} character having the same
symmetry as the bare, unrenormalized orbital. Examining
Fig.~\ref{ED_df}, we see that the central Ti site shows the expected
$d_{xy}$ character, while the {\it tail} of the orbital sitting at the
O and Cl sites has appreciable weight shaped to O-$p_{x}$,$p_{y}$ and
Cl-$p_{x}$,$p_{y}$ symmetries which indicates appreciable
renormalization effect coming from O-$p_{x}$,$p_{y}$ and
Cl-$p_{x}$,$p_{y}$
%\cite{explain2}.
%% as was already evident from hopping integrals \cite{explain2}.  
This analysis leads us to conclude that the
equilibrium state of TiOCl at T $>$ T$_{c2}$  is given by the Ti $d_{xy}$-Ti
$d_{xy}$ direct coupling with {\it sizable} interaction via the
surrounding O and Cl.

\putfigc 
{\it Distorted structures -}
A fascinating feature about this system is its anomalous
behavior\cite{Kataev_03,lemmens_03_2} near the phase transition at
T$_{c2}$  which has been attributed to competing lattice,
spin and orbital degrees of freedom.
As described above, {\it ab-initio} calculations on the crystal
structure measured at high temperatures
T$>$T$_{c2}$ predict unambiguously an electronic
groundstate formed of non-degenerate Ti 3$d_{xy}$ orbitals while 
 $d_{yz}$ and $d_{xz}$ are higher in
energy.  However, displacements of the Ti sites due to phonon modes are
expected to influence the electronic state and it remains to be seen
whether
during vibration occupation of certain $d$ orbitals changes drastically.
This would indicate that orbital fluctuations are important in this system
since small atomic displacements  can induce a change in the electronic 
occupation. 
%In TiOCl one should then observe a change of
%occupation from the $d_{xy}$ orbital
%to  the orbitals $d_{xz}$ and $d_{yz}$.
%leading to the idea of orbital ordering in
%this system which would be described by alternating $d_{yz}$ and
%$d_{xz}$ along $a$.

In order to investigate this scenario, we have performed  a
frozen phonon study where we calculated the groundstate properties
for distorted structures according to the various Raman active A$_g$
phonon modes expected for the $Pmmn$ space group.   These modes
were calculated by considering a shell model \cite{natasha, shell} for
TiOCl.  The displacements considered in the {\it ab initio} calculation
were chosen according to the eigenvector components of the A$_g$
modes obtained from the shell calculation.
The A$_g$ phonon modes define ion displacements along the $c$
axis. Raman observations\cite{lemmens_03_2} suggest that the Ti-Cl
in-phase  zone boundary  A$_g$ mode plays an important role above the
transition at T$_{c2}$.  From those observations 
it is not conclusive \cite{lemmens_p} whether it corresponds to a
 phonon for a unit-cell doubled along $c$, $b$ or $a$. The doubling of the
unit cell along $b$ reported by NMR \cite{Imai_03} happens in the gapped region well below
T$_{c2}$.
In view of  the above,  we
calculated the groundstate properties of various cases where the ions
were displaced from their equilibrium
 position by considering displacements in magnitude  smaller than 4 $\%$
of the lattice constant according to various
A$_g$ phonon modes\cite{lemmens_03_2,natasha}.  We have considered three
types of
Ti-Cl in-phase
distortions as shown in Fig.\ ~\ref{distort}. Case -(a) No doubling
 of the unit cell.
  Case -(b):  The
unit cell doubled  along $b$. In this case, the distorted
structure is monoclinic with two  crystallographically
inequivalent Ti atoms. Case
-(c): The  
unit cell doubled along $c$.  Here the unit cell remains orthorhombic 
with two inequivalent Ti ions, Ti1 and Ti2.
\putfigd
 We have carried
out LDA(GGA) and LDA+U calculations in FM spin configurations for each of 
these  cases.
 In case -(a) as drawn in Fig.\ \ref{distort} a)
the groundstate remains the same as for the undistorted
case with $d_{xy}$ as the lowest-energy, occupied orbital. However in
the
distorted structure with the atoms displaced in the opposite sense
to Fig.\ \ref{distort} a), we observe  a change in the
orbital occupancies. The groundstate  is described by the 
 $d_{yz}$ and $d_{xz}$ orbitals {\it instead of $d_{xy}$}.  In case -(b) 
%lifts the near degeneracy 
the distortion enlarges
 the splitting between
$d_{yz}$ and $d_{xz}$ while the groundstate remains described by the $d_{xy}$
orbital. In case -(c) we observe 
  a  change in the orbital occupancies for Ti1 atoms as
is evident from the LDA+U bandstructure results shown in
Fig.~\ref{distorsion_bands}. This may be compared
with the change of orbital occupancies for case (a) with atoms moved in
opposite direction than drawn in Fig.\ \ref{distort} a).
The groundstate in case -(c) is described by the 
 $d_{yz}$ and $d_{xz}$ orbitals  at Ti1 while
$d_{xy}$ remains
as the lowest-energy, occupied orbital at Ti2.  
\putfige
These results support therefore the concept of possible orbital fluctuations 
in this system induced by strong electron-phonon interactions.

Although we have considered only a few chosen cases of distorted
structures, and have certainly not explored all the possible allowed
distortions in this material, the change of groundstate in some
of the above distorted calculations
undoubtedly suggests that lattice, charge and orbital degrees of
freedom are intimately related in this system. 
 One may argue about the size of the distortion considered.  Further
 analysis with smaller distortions shows that this effect is still
 observed \cite{new}. 
 In conclusion, this
system is not orbital ordered but   may be subject to  orbital
fluctuations at high temperatures.  We hope that our study
will stimulate further investigations in the direction of understanding
this fascinating compound.

\vspace{0.2cm}

{\it Acknowledgments}

 We  acknowledge fruitful discussions with P. Lemmens,
 D. Khomskii, N. Kovaleva and M. Gr\"uninger and
 thank the Deutsche Forschungsgemeinschaft for financial support.
  
%%%%%%%%%%%%%%%%%%%%%%%%%%%%%%%%%%%%%%%%%%%%%%%%%%%%%%%%%%%

%%%555
%%%555

%%%%%%%%%%%%%%%%%%%%%%%%%%%%%%%%%%%%%%%%%%%%%%%%%%%%%%%%%%%%
%%%%%%%%%%%%%%%%%%%%%%%%%%%%%%%%%%%%%%%%%%%%%%%%%%%%%%%%%%%%

%%%%%%%%%%%%%%%%%%%%%%%%%%%%%%%%%%%%%%%%%%%%%%%%%%%%%%%%%%%%%%%%%%%%%%%%%
%%%%%%%%%%%%%%%%%%%%%%%%%%%%%%%%%%%%%%%%%

%\newpage

%\begin{onecolumn}

%%%%%%%%%%%%%%%%%%%%%%%%%%%%%%%%%%%%%%%%%%%%%%%%%%%%%%%%%%%%%%


\begin{thebibliography}{99}

\bibitem{review} see for a review P. Lemmens, C. Gros,
G. G\"untherodt, Physics Reports {\bf 375}, 1 (2003).

%\bibitem{chitov} G. Y. Chitov, C. Gros, cond-mat/0310494

%\bibitem{Isobe_02} M. Isobe, E. Ninomiya, A. V. Vasilev, and Y. Ueda,
%Jour. Phys. Soc. Jap. {\bf 71}, 1423 (2002).

%\bibitem{Konst_03} M. J. Konstantinovic, J. van den Brink,
%Z. V. Popovic, V. V. Moshchalkov, M. Isobe, and Y. Ueda,
%cond-mat/0210191

\bibitem{Keimer_00} B. Keimer {\it et al.}, Phys. Rev. Lett. {\bf 85},
3946 (2000).

\bibitem{Khaliullin_00} G. Khaliullin and S. Maekawa,
Phys. Rev. Lett. {\bf 85}, 3950 (2000).

\bibitem{Cwick_03} M. Cwick {\it et al.}, Phys. Rev. B {\bf 68},
060401 (2003).

\bibitem{Pavarini_03} E. Pavarini {\it et al.}
%, S. Biermann, A. Poteryaev,
%A. I. Lichtenstein, A. Georges, and O. K. Andersen, 
cond-mat/0309102

\bibitem{Craco_03} L. Craco, M. S. Laad, S. Leoni, and
E. M\"uller-Hartmann, cond-mat/0309370

\bibitem{Seidel_03} A. Seidel {\it et al.}
%, C. A. Marianetti, F. C. Chou, G. Ceder,
%and P. A. Lee, 
Phys. Rev. B {\bf 67}, 020405 (2003).

\bibitem{Imai_03} T. Imai and F.C. Chou, cond-mat/0301425



\bibitem{Kataev_03} V. Kataev, J. Baier, A. M\"oller, L. Jongen,
G. Meyer, and A. Freimuth, cond-mat/0305317.

\bibitem{lemmens_03_2} P. Lemmens, K. Y. Choi, G. Caimi, L.Degiorgi,
N. N. Kovaleva, A. Seidel, and F.C. Chou, cond-mat/0307502

\bibitem{Andersen_75} O. K. Andersen, Phys. Rev. B {\bf 12}, 3060
(1975).

\bibitem{newlmto} O.K. Andersen and T. Saha-Dasgupta, Phys. Rev. {\bf
B62}, R16219 (2000) and references therein.

\bibitem{explain} The site symmetry of the Ti ion
in TiOCl is $C_{2v}$. With the given crystal structure and
symmetry at the Ti site, the $d_{xz}$ and $d_{yz}$ orbitals are
not  degenerate. They show a  small
crystal field splitting of  about  one-tenth of an eV.

\bibitem{Perdew_96} J.P. Perdew, K. Burke, and M. Ernzerhof,
Phys. Rev.  Lett. {\bf 77}, 3865 (1996).

\bibitem{Anisimov_97} V. Anisimov, F. Aryasetiawan, and
A. I. Lichtenstein, J. Phys.: Condens. Matter {\bf 9}, 767 (1997).

\bibitem{Koepernik_99} K. Koepernik and H. Eschrig, Phys. Rev. B {\bf
59}, 1743 (1999).


\bibitem{convergence}
The calculations were checked carefully for convergency with respect to
the basis set and the k-mesh (4800 k-points have been used in the
Brillouin zone of the simple cell).

\bibitem{ref_frame}
Note that this new coordinate system is not exactly the local
reference system for the Ti ion. In this chosen reference
frame the $d_{xz}$ and the $d_{yz}$ orbitals are  degenerate.

%\bibitem{Elfimov_03} S. Elfimov, T. Saha-Dasgupta and M. A. Korotin,
%Phys. Rev. B {\bf 68}, 114105 (2003).

\bibitem{Valenti_03_t} R. Valent\'\i, T. Saha-Dasgupta, C. Gros,  H. Rosner,
Phys. Rev. B  {\bf 67}, 245110 (2003).

%\bibitem{explain2} We see also traces of other
%integrated out $d$-characters at neighboring Ti sites born via  small
%mixing of e$_g$ and t$_{2g}$ characters due to deviation of the angle
% $\alpha$ from 180$^{o}$.

\bibitem{natasha} Data provided by N. N. Kovaleva, MPI Stuttgart
    (unpublished).

\bibitem{shell} J. W. Powell {\it et al.}
%, G. S. Edwards, L. Genzel, F. Kremer,  A. Wittlin,
%                W. Kubasek and W. Peticolas, 
Phys. Rev. A {\bf 35},
    3929 (1987).

\bibitem{lemmens_p} P. Lemmens, private communication.

\bibitem{new}   L. Pisani, T. Saha-Dasgupta, N. N. Kovaleva, R. Valent\'\i,
              in preparation.

%%%%%%%%%%%%%%%%%%%%%%%%%%%%%%%%%%%%%%%%%%%%%%%%%%%%%
%%%%%%%%%%%%%%%%%%%%%%%%%%%%%%%%%%%%%%%%%%%%%%%%%%%%%%

\end{thebibliography}
\end{document}